\documentclass[11pt]{article}

\usepackage[preprint]{acl}

\usepackage{times}
\usepackage{latexsym}

\usepackage[T1]{fontenc}
\usepackage[utf8]{inputenc}

\usepackage{microtype}
\usepackage{amsmath}
\usepackage{multirow}
\usepackage{booktabs}
\usepackage{subcaption}
\usepackage{inconsolata}

\usepackage{graphicx}

\title{Learning from Distribution: Rethinking One-Hot Supervision for ASR and QA in Speech Language Models}

\author{
    \textbf{Yujie Guo}\textsuperscript{1, $\ast$},
    \textbf{Hongjie Chen}\textsuperscript{2},
    \textbf{Jiaming Zhou}\textsuperscript{1},
    \textbf{Jian Kang}\textsuperscript{2},
    \textbf{Jie Li}\textsuperscript{2, $\dagger$},
    \textbf{Yongxiang Li}\textsuperscript{2},
    \textbf{Yong Qin}\textsuperscript{1, $\dagger$}
    \\
    \textsuperscript{1}College of Computer Science, Nankai University
    \\
    \textsuperscript{2}Xingchen AGI Lab, China Telecom Artificial Intelligence Technology Co. Ltd
    \\
    \small{guoyujie02@mail.nankai.edu.cn, chenhj37@chinatelecom.cn, qinyong@nankai.edu.cn}
}
\begin{document}
\maketitle

\begingroup
\renewcommand{\thefootnote}{}
\footnotetext{$^\ast$ This work was done during an internship at Xingchen AGI Lab.}
\footnotetext{$^\dagger$ Yong Qin and Jie Li are corresponding authors.}
\endgroup

\begin{abstract}
Speech language models (SLMs) exhibit strong capabilities in automatic speech recognition (ASR) and general-purpose speech understanding, yet ASR-specific training often degrades spoken question answering (QA) performance. We revisit this conflict from the perspective of task-specific input--output mappings and one-hot supervision. Our findings reveal pronounced task-asymmetric effects: ASR-only training severely limits QA capability, whereas QA-only training retains reasonable ASR performance. We hypothesize that the relatively deterministic mapping of ASR, combined with single-reference supervision, biases models toward transcription-specific behavior. Motivated by this perspective, we advocate broadening ASR learning objectives while preserving recognition accuracy. We propose Learning from Distribution (LFD), which aligns speech-conditioned predictions with the conditional output distributions of a pretrained text LLM. Extensive experiments demonstrate that LFD improves both ASR and QA performance, exhibits superior data efficiency, and remains effective under full-parameter fine-tuning. CoT-ASR further supports the broader applicability of our proposed supervision principle.
\end{abstract}

\section{Introduction}

Recent advances in large language models (LLMs) have accelerated the development of speech language models (SLMs), improving both automatic speech recognition (ASR) and general-purpose speech understanding~\cite{qwen2audio,Qwen3-omni,stepaudio2.5,salmonn2,DeSTA2}. Unlike conventional ASR systems, SLMs enable instruction-conditioned recognition, incorporating contextual information and domain-specific hotwords into transcription~\cite{zheng2026multilevel,contextasr-bench}. By leveraging pretrained LLM capabilities, SLMs offer a framework for unifying speech recognition and general-purpose language intelligence~\cite{Blsp-kd,wang2024freeze}.
% Recent advances in large language models (LLMs) have accelerated the development of speech language models (SLMs), improving both automatic speech recognition (ASR) and general-purpose speech understanding~\cite{qwen2audio,Qwen3-omni,stepaudio2.5,salmonn2}. Unlike conventional ASR systems, SLMs enable instruction-conditioned recognition, incorporating contextual information and domain-specific hotwords into transcription~\cite{zheng2026multilevel}. By leveraging pretrained LLM capabilities, SLMs offer a framework for unifying speech recognition and general-purpose language intelligence.

Despite recent advances, general-purpose SLMs still lag behind specialized ASR systems in recognition accuracy and robustness, motivating ASR-specific fine-tuning~\cite{qwen3asr,Marco-ASR}. However, ASR specialization may compromise general-purpose capabilities, particularly spoken question answering (QA), by biasing models toward transcription~\cite{Diva,hsiao2025analyzing,Salad,ts-opd}. This raises a key challenge: improving ASR without sacrificing broader language capabilities.

% Despite these advances, general-purpose SLMs still face challenges in matching the recognition accuracy and robustness of specialized ASR systems, motivating ASR-specific fine-tuning~\cite{qwen3asr,Marco-ASR}. However, improving ASR performance may compromise general-purpose capabilities, particularly spoken question answering (QA), as models become increasingly biased toward transcription~\cite{Diva,hsiao2025analyzing}. This creates a fundamental tension between ASR specialization and general-purpose speech understanding, raising the challenge of improving recognition without sacrificing broader capabilities.

Existing studies have primarily attributed the degradation of general-purpose capabilities in SLMs to catastrophic forgetting and cross-modal misalignment~\cite{hsiao2025analyzing,Diva,Salad,DeSTA2,tan2025ssr}. To mitigate these issues, various strategies have been explored, including experience replay, parameter-efficient adaptation, model merging, and cross-modal knowledge distillation~\cite{hsiao2025analyzing,Blsp-kd,alignformer,wang2026cross,ts-opd,xopd}. Nonetheless, how task-specific data characteristics and supervision objectives contribute to capability degradation remains insufficiently understood.
% Existing studies have primarily examined the degradation of general-purpose capabilities in SLMs from the perspectives of catastrophic forgetting and cross-modal misalignment~\cite{hsiao2025analyzing,Diva,Salad}. Catastrophic forgetting refers to the loss of pretrained capabilities during task-specific adaptation~\cite{hsiao2025analyzing,eeckt2022weight}, while cross-modal misalignment limits the ability of speech inputs to leverage the pretrained LLM's language understanding and generation capabilities~\cite{Blsp-kd,alignformer,tan2025ssr}. To address these issues, previous research has explored strategies such as experience replay, parameter-efficient adaptation, model merging, and cross-modal knowledge distillation~\cite{hsiao2025analyzing,Blsp-kd,Diva,Salad,wang2026cross}. Although these approaches have shown promise in preserving general-purpose capabilities, a fundamental question remains: how do the characteristics of task-specific training data and their supervision objectives contribute to capability degradation?

Addressing this question requires examining the relationship between task-specific mappings and training supervision. Although ASR and QA share the same generative formulation, ASR follows an approximately one-to-one mapping, whereas QA admits multiple valid responses. Despite these differences, both tasks commonly rely on one-hot cross-entropy (CE) supervision, which reinforces a single reference output per training example. While CE naturally aligns with the deterministic mapping of ASR, diverse QA responses can collectively approximate the underlying output distribution. We argue that these distinct mappings may lead to asymmetric learning behaviors under the same One-hot supervision objective.

% Addressing this question calls for a closer examination of the relationship between task characteristics and training supervision. Although ASR and spoken QA can both be formulated as speech-conditioned text generation, they exhibit distinct input--output mappings. ASR typically follows an approximately one-to-one mapping from speech to transcription, whereas QA admits a one-to-many mapping with multiple semantically valid responses. Despite these differences, both tasks are commonly optimized using cross-entropy (CE) loss, which reinforces a single reference output for each training example. Such supervision naturally aligns with the relatively deterministic mapping of ASR. For QA, although each training example provides only one reference response, CE can still approximate the underlying conditional output distribution given sufficiently diverse and representative training data. Thus, while CE is applicable to both tasks, their distinct mapping characteristics may lead to different learning behaviors under the same supervision objective.

We hypothesize that the interaction between task-specific input--output mappings and one-hot supervision contributes to the ASR--QA conflict. To address this, we advocate broadening ASR learning objectives beyond single-reference targets. We propose Learning from Distribution (LFD), which realizes this principle by aligning speech-conditioned predictions with the conditional output distributions of a pretrained text LLM. We further explore an alternative training strategy to demonstrate the broader applicability of this principle. Our main contributions are summarized as follows:

% We hypothesize that the interaction between task-specific input--output mappings and one-hot supervision contributes to the ASR--QA conflict. To mitigate this conflict, we advocate moving beyond single-reference supervision to preserve the broader generation capabilities of pretrained LLMs. Following this perspective, we propose Learning from Distribution (LFD), which aligns speech-conditioned predictions with the conditional output distributions of a pretrained text LLM. We further validate this perspective through an alternative training strategy, suggesting that its benefits extend beyond LFD. Our main contributions are summarized as follows:

\begin{itemize}
    \item We investigate the ASR--QA conflict from a new perspective by examining the interaction between task-specific input--output mappings and one-hot supervision. Through systematic comparisons of ASR-only and QA-only training, we reveal pronounced task-asymmetric effects and provide a mapping-based explanation for this phenomenon.

    \item We propose broadening ASR learning objectives based on the distinct input--output mappings of ASR and QA. We introduce Learning from Distribution (LFD) as one realization of this principle and further demonstrate its generality through Chain-of-Thought ASR.

    \item Extensive experiments demonstrate that LFD improves both ASR and QA performance with superior data efficiency. Tail-aware top-$k$ supervision maintains competitive performance while enabling compact storage of teacher distributions and efficient offline supervision.

    %\item We provide a new perspective on the conflict between ASR specialization and general-purpose capabilities by examining the interaction between task-specific input--output mappings and one-hot supervision. Through systematic comparisons of ASR- and QA-oriented training, we reveal pronounced task-asymmetric effects and offer an explanation grounded in their distinct mapping characteristics.

    % \item We propose a mapping-oriented perspective for mitigating the conflict between ASR and general-purpose capabilities. Based on the distinct input--output mappings of ASR and QA, we introduce LFD as one realization of this perspective and further validate the proposed principle through an alternative training strategy, demonstrating that its effectiveness extends beyond a single method.
    
    % \item We demonstrate the effectiveness and practicality of LFD through extensive experiments. LFD substantially improves both ASR and QA performance, exhibits superior data efficiency, and achieves competitive results with tail-aware top-$k$ supervision, enabling compact storage of teacher distributions and efficient offline supervision.
\end{itemize}

\section{A Mapping Perspective on ASR--QA Conflict}
\label{sec:mapping}
ASR and QA can both be formulated as speech-conditioned text generation:
\begin{equation}
    p_{\phi}(Y \mid X, I)
    = \prod_{t=1}^{T}
    p_{\phi}(y_t \mid X, I, y_{<t}),
\end{equation}
where $X$ is the speech input, $I$ is the task instruction, $Y$ is the output sequence, and $\phi$ denotes the trainable adapter parameters.

\begin{figure}[t]
    \centering
    \includegraphics[width=\columnwidth]{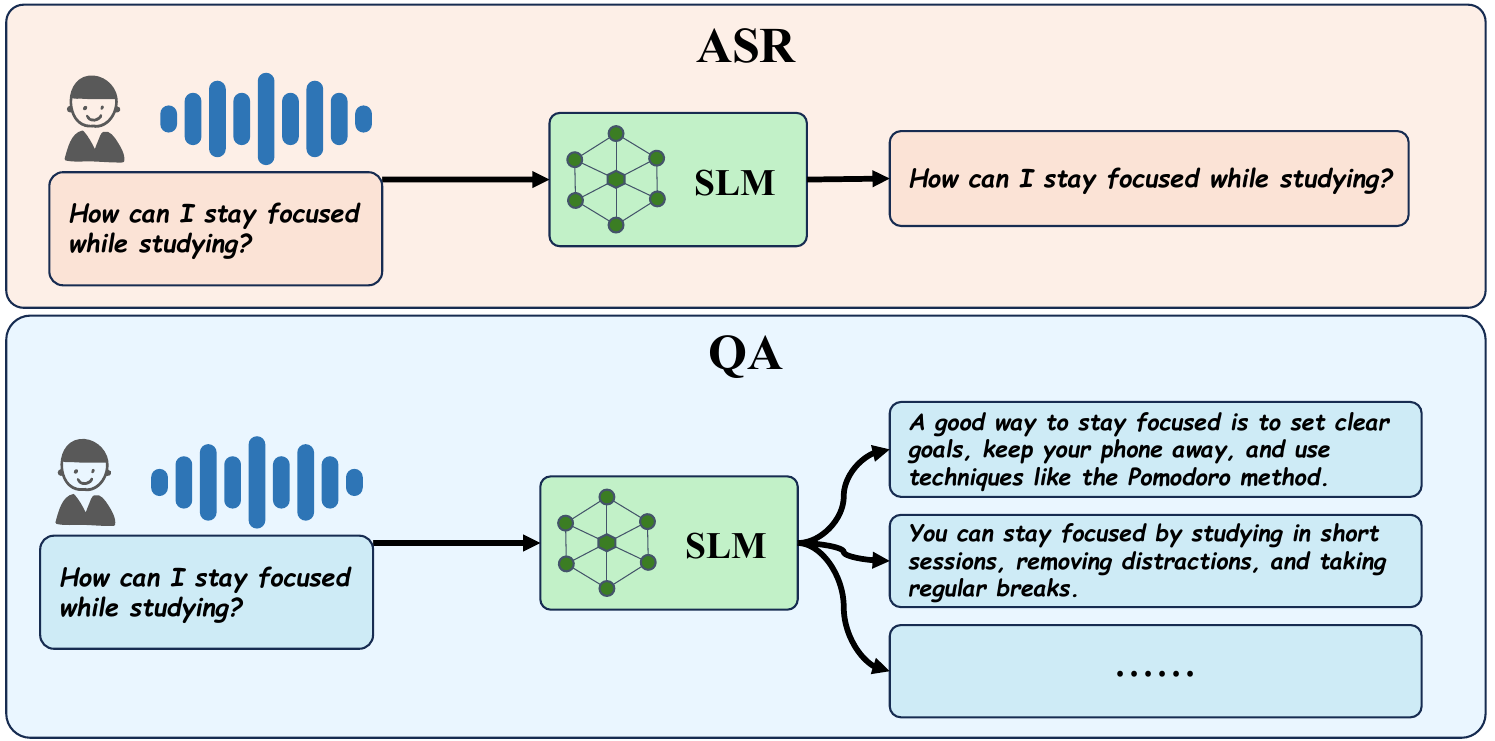}
    \caption{Illustration of ASR's approximately one-to-one mapping and QA's one-to-many mapping with multiple valid responses.}
    \label{fig:mapping}
\end{figure}

\begin{figure*}[t]
  \centering
  \includegraphics[width=1.0\textwidth]{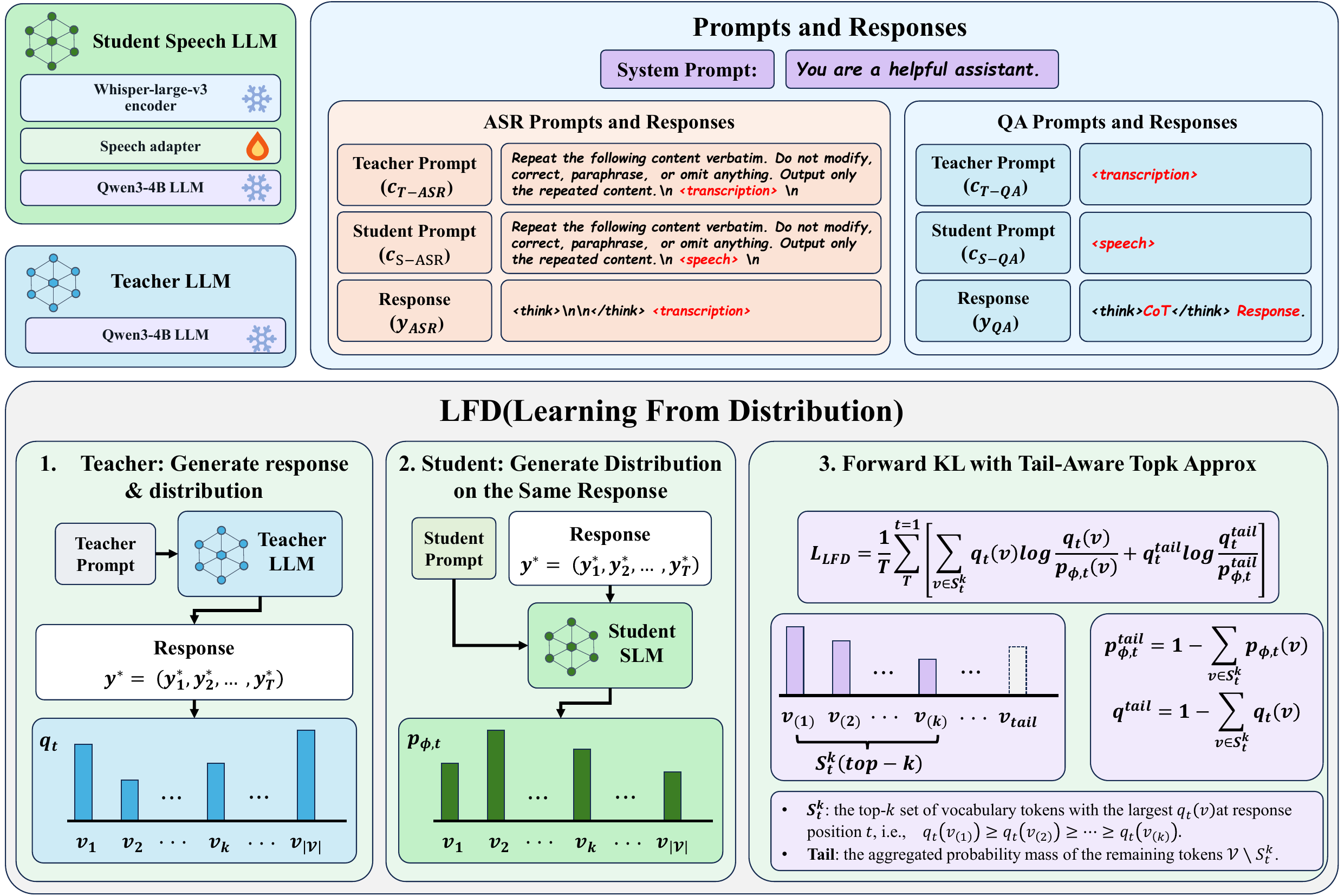}
\caption{Overview of LFD. The text LLM teacher generates responses and token distributions, which the Speech LLM student learns via forward KL divergence with a tail-aware top-$k$ approximation.}
\label{fig:lfd}
\end{figure*}

Despite sharing the same formulation, ASR and QA exhibit distinct input--output mappings, as illustrated in Fig.~\ref{fig:mapping}. ASR typically involves an approximately one-to-one mapping from speech to transcription, whereas QA admits a one-to-many mapping with multiple semantically valid responses. These mapping differences are reflected in the entropy characteristics of their conditional output distributions. ASR generally exhibits lower conditional output entropy due to its relatively deterministic transcription targets, whereas QA tends to exhibit higher entropy owing to the diversity of valid responses. We investigate these task-dependent entropy characteristics in SLMs and text LLMs, with the main results presented in Section~\ref{sec:distribution_analysis} and additional analyses provided in Appendix~\ref{sec:appendix_entropy}.
% Despite sharing the same formulation, ASR and QA exhibit distinct input--output mappings, as illustrated in Fig.~\ref{fig:mapping}. ASR typically involves an approximately one-to-one mapping from speech to transcription, whereas QA admits a one-to-many mapping with multiple semantically valid responses. These mapping differences are reflected in the entropy characteristics of their conditional output distributions. ASR generally exhibits lower conditional output entropy due to its relatively deterministic transcription targets, whereas QA tends to exhibit higher entropy owing to the diversity of valid responses. We further investigate these task-dependent entropy characteristics in both SLMs and text LLMs, with the main results presented in Section~\ref{sec:distribution_analysis} and additional analyses provided in Appendix~\ref{sec:appendix_entropy}.

Conventional cross-entropy (CE) applies the same one-hot supervision to both tasks:
\begin{equation}
    \mathcal{L}_{\mathrm{CE}}
    = -\sum_{t=1}^{T}
    \log p_{\phi}(y_t^* \mid X, I, y_{<t}^*),
\end{equation}
where $y_t^*$ denotes the reference token. For each training example, CE reinforces the conditional probability of a single reference output. For ASR, this supervision naturally aligns with its relatively deterministic input--output mapping, consistently reinforcing speech-to-transcription alignment. For QA, although each training example provides only one reference response, the inherent diversity of valid responses can be captured across sufficiently large training datasets, allowing CE to approximate the underlying conditional output distribution.

Nevertheless, the distinct mapping characteristics of ASR and QA may lead to different learning behaviors under the same supervision objective. In particular, repeated optimization toward deterministic ASR targets may bias speech--language alignment toward transcription-specific behavior, limiting access to the pretrained LLM's broader generation capabilities. In contrast, diverse QA supervision may expose the model to richer generation patterns, resulting in different cross-task effects. We therefore hypothesize that the ASR--QA conflict is partly shaped by the interaction between task-specific input--output mappings and one-hot supervision, rather than being an inherent consequence of ASR data alone.

\section{Method}
\label{sec:method}

\subsection{Rethinking ASR Supervision}

As discussed in Section~\ref{sec:mapping}, the approximately one-to-one mapping of ASR naturally aligns with one-hot CE supervision. However, repeatedly optimizing deterministic transcription targets may bias speech--language alignment toward transcription-specific behavior, limiting the SLM's ability to leverage the pretrained LLM's broader generation capabilities. This suggests that the ASR--QA conflict may be mitigated by reconsidering how ASR data is used for supervision.

Motivated by this observation, we advocate broadening ASR learning objectives beyond single-reference supervision. Rather than relying solely on deterministic transcription targets, ASR training should incorporate richer supervision that preserves the pretrained LLM's broader generation behavior. This principle aims to maintain accurate speech recognition while reducing the risk of compromising general-purpose capabilities.

% Motivated by this observation, we advocate broadening the learning objective of ASR beyond single-reference supervision. Rather than learning exclusively from deterministic transcription targets, ASR training should incorporate richer supervisory information that preserves the broader generation behavior of the pretrained LLM. This principle aims to maintain accurate speech recognition while reducing the risk of compromising general-purpose capabilities.

\subsection{Learning from Distribution}

Following this principle, we propose Learning from Distribution (LFD), which broadens the learning objective by aligning speech-conditioned predictions with the conditional output distributions of a pretrained text LLM. Instead of relying solely on one-hot reference tokens, LFD preserves the LLM's probability allocation across possible predictions, providing richer supervision for ASR while aiming to retain its broader generation capabilities. As illustrated in Fig.~\ref{fig:lfd}, a frozen text LLM teacher and an SLM student are conditioned on task-specific text and speech inputs, respectively.

Given the target response $Y^*$, teacher forcing is applied to both models along the same response trajectory. Their token distributions are defined as
\begin{equation}
    \begin{aligned}
    q_t &= p_T(\cdot \mid C_{\mathrm{text}}, y_{<t}^*), \\
    p_{\phi,t} &= p_{\phi}(\cdot \mid X, I, y_{<t}^*),
    \end{aligned}
\end{equation}
where $C_{\mathrm{text}}$ denotes the task-conditioned textual context and $p_T$ the teacher distribution. LFD minimizes the forward KL divergence between them:
\begin{equation}
    \mathcal{L}_{\mathrm{full}}
    = \frac{1}{T}\sum_{t=1}^{T}
    D_{\mathrm{KL}}(q_t \| p_{\phi,t}).
\end{equation}
This objective transfers the LLM's conditional generation behavior to speech, rather than supervising individual reference tokens alone.

\subsection{Tail-Aware Top-k Distribution Learning}

Directly storing full-vocabulary teacher distributions incurs substantial storage overhead~\cite{peng2025pre,zaidi2025sparse,ryskulov2026efficient}. To enable efficient distribution supervision, we retain only the top-$k$ tokens with the highest teacher probabilities~\cite{li2025bild}. Let $\mathcal{S}_t^k$ denote the selected token set at position $t$.

However, naively truncating the teacher distribution yields
\begin{equation}
    \mathcal{L}_{\mathrm{trunc}}
    = -\sum_{v \in \mathcal{S}_t^k}
    q_t(v)\log p_{\phi,t}(v),
\end{equation}
which encourages the student to concentrate its probability mass on the selected tokens. In particular, when $k=1$, this objective reduces to a scaled cross-entropy loss on the teacher's most probable token, losing the intended benefits of distributional supervision.

Motivated by the importance of preserving tail probability information~\cite{zaidi2025sparse,huang2026tail,meng2026unbiased}, we adopt a tail-aware KL objective that retains the aggregate probability mass outside the top-$k$ set. The teacher and student tail probabilities are defined as
\begin{equation}
    \begin{aligned}
    q_t^{\mathrm{tail}}
    &= 1-\sum_{v\in\mathcal{S}_t^k}q_t(v),\\
    p_{\phi,t}^{\mathrm{tail}}
    &= 1-\sum_{v\in\mathcal{S}_t^k}p_{\phi,t}(v).
    \end{aligned}
\end{equation}

The resulting objective is
\begin{equation}
    \begin{aligned}
    \mathcal{L}_{\mathrm{LFD}}
    = \frac{1}{T}\sum_{t=1}^{T}\Bigg[
    &\sum_{v\in\mathcal{S}_t^k}
    q_t(v)\log\frac{q_t(v)}{p_{\phi,t}(v)}\\
    &+q_t^{\mathrm{tail}}
    \log\frac{q_t^{\mathrm{tail}}}
             {p_{\phi,t}^{\mathrm{tail}}}
    \Bigg].
    \end{aligned}
\end{equation}

By preserving residual probability mass, this objective constrains the student's predictions within and beyond the top-$k$ set, enabling efficient distribution matching without storing full-vocabulary teacher distributions.
%By explicitly preserving the residual probability mass, this objective constrains the student's probability allocation both within and outside the selected top-$k$ set. It provides an efficient approximation to full-vocabulary distribution matching without explicitly storing the entire teacher distribution.

\section{Experimental Setup}

\subsection{Model Architecture and Training Setup}
\label{sec:experimental_setup}
Our SLM consists of a pretrained Whisper-large-v3 speech encoder~\cite{whisper}, a modality adapter, and a Qwen3-4B LLM\footnote{https://huggingface.co/Qwen/Qwen3-4B}~\cite{qwen3}. We employ Qwen3-4B as the frozen text teacher for distribution supervision.

We use 8,000 hours of speech from Emilia~\cite{emilia} to construct ASR and QA training data. ASR uses the original transcripts, while QA data is generated from transcripts using Qwen3-4B. We evaluate four settings: CE-ASR, CE-QA, LFD-ASR, and LFD-QA. Within each task, CE and LFD share identical data, architectures, and initializations, differing only in supervision objectives. LFD uses $k=256$ by default.
% We use 8,000 hours of speech from the Emilia corpus~\cite{emilia} to construct ASR and QA training data. For ASR, the original transcripts serve as supervision targets. For QA, we use Qwen3-4B to generate question--answer data based on the speech transcripts. We investigate four training settings: CE-ASR, CE-QA, LFD-ASR, and LFD-QA. Within each task setting, CE and LFD use identical training data, model architectures, and initializations, differing only in their supervision objectives. LFD adopts a default top-$k$ value of 256.

In the main experiments, we focus on speech--language alignment, where the pretrained speech encoder and LLM remain frozen, and only the modality adapter is optimized. All models are trained for one epoch with a learning rate of $1\times10^{-4}$ and a cosine learning rate scheduler.

\subsection{Evaluation Benchmarks and Metrics}

We evaluate ASR performance on the LibriSpeech test-clean and test-other sets~\cite{librispeech} using word error rate (WER). For spoken QA, we adopt the TELEVAL benchmark~\cite{televal}, including LlamaQA-en, TriviaQA-en, and WebQ-en, and report answer accuracy (ACC). Detailed descriptions of the benchmarks and evaluation metrics are provided in Appendix~\ref{sec:appendix_evaluation}.

\begin{table*}[t]
\caption{ASR and QA performance under CE and LFD supervision
with ASR-only and QA-only training data.
ASR performance is measured by WER(\%),
while QA performance is evaluated by ACC(\%) on TELEVAL.
}
\label{tab:mainres}
\renewcommand{\arraystretch}{1.05}
\centering
\resizebox{\textwidth}{!}
{%
\begin{tabular}{cccccccccc}
\hline
\multirow{2}{*}{System} & \multirow{2}{*}{\begin{tabular}[c]{@{}c@{}}Training\\ Strategy\end{tabular}} & \multirow{2}{*}{\begin{tabular}[c]{@{}c@{}}Training\\ Data\end{tabular}} & \multicolumn{3}{c}{ASR(Librispeech $\downarrow$)} & \multicolumn{4}{c}{QA(TELEVAL $\uparrow$)}                        \\
                        &                                                                              &                                                                          & clean           & other          & Overall        & LlamaQA-en     & TriviaQA-en    & WebQ-en        & Overall        \\  \midrule
A1                      & CE                                                                           & \multirow{2}{*}{ASR}                                                     & 4.83            & 8.96           & 6.89           & 1.67                                                  & 2.87                                                   & 2.37                                               & 2.44           \\
A2                      & LFD                                                                          &                                                                          & \textbf{4.51}   & \textbf{8.63}  & \textbf{6.57}  & \textbf{65.33}                                        & \textbf{31.78}                                         & \textbf{39.94}                                     & \textbf{40.20} \\ \midrule
B1                      & CE                                                                           & \multirow{2}{*}{QA}                                                      & 6.35            & 10.11          & 8.23           & 72.00                                                 & 36.08                                                  & 42.05                                              & 43.35          \\
B2                      & LFD                                                                          &                                                                          & \textbf{5.18}   & \textbf{9.23}  & \textbf{7.20}  & \textbf{73.33}                                        & \textbf{36.56}                                         & \textbf{42.57}                                     & \textbf{43.93} \\ \bottomrule
\end{tabular}%
}
\end{table*}

\section{Results and Analysis}

\subsection{Task-Asymmetric Effects of One-Hot Supervision}

Table~\ref{tab:mainres} compares CE and LFD under ASR-only and QA-only training settings.

\textbf{One-hot supervision exhibits task-asymmetric effects consistent with distinct input--output mappings.}
Comparing A1 and B1, ASR-only CE achieves strong recognition performance but severely limited QA capability. In contrast, QA-only CE maintains strong QA performance while retaining reasonable ASR capability. This pronounced asymmetry is consistent with our mapping perspective: the approximately one-to-one mapping of ASR, combined with one-hot supervision, repeatedly reinforces deterministic transcription targets, potentially biasing the adapter toward transcription-specific alignment. In contrast, the inherent diversity of QA responses may expose the model to richer generation patterns, resulting in different cross-task behaviors.

\textbf{Broadening ASR supervision mitigates the degradation of general-purpose capabilities.}
Comparing A1 and A2, replacing CE with LFD under the same ASR-only training setting substantially improves QA capability while further enhancing ASR performance. Notably, these improvements are achieved without introducing any QA training data. By broadening the learning objective beyond individual reference tokens, LFD enables the model to better leverage the pretrained LLM's broader generation capabilities. This finding suggests that the degradation associated with ASR specialization depends not only on the training data itself, but also on how the model learns from it.

\textbf{Distributional supervision mitigates task-asymmetric learning effects.}
Comparing B1 and B2, LFD improves both ASR and QA under QA-only training. Comparing A2 and B2, LFD substantially narrows the cross-task gap between ASR-only and QA-only training, while each retains an advantage on its primary task. These findings suggest that learning richer conditional output distributions mitigates task-asymmetric effects, supporting our principle of broadening supervision beyond single-reference targets.

% \textbf{Distributional supervision mitigates task-asymmetric learning effects.} Comparing B1 and B2, LFD also improves both ASR and QA performance under QA-only training. Furthermore, comparing A2 and B2, the cross-task performance gap between ASR-only and QA-only training is substantially reduced under LFD, although each setting retains an advantage on its primary task. These observations suggest that learning richer conditional output distributions can alleviate the asymmetric effects associated with different task mappings, further supporting our proposed principle of broadening supervision beyond single-reference targets.

\begin{table}[t]
\centering
\caption{Scaling results of CE and LFD under ASR-only and QA-only training settings with different amounts of training data.}
\label{tab:scaling_overall}
\renewcommand{\arraystretch}{1.05}
\setlength{\tabcolsep}{4pt}
\small
\resizebox{\columnwidth}{!}{%
\begin{tabular}{llcccc}
\toprule
\multirow{2}{*}{\begin{tabular}[c]{@{}c@{}}Training\\ Data\end{tabular}}
& \multirow{2}{*}{\begin{tabular}[c]{@{}c@{}}Training\\ Strategy\end{tabular}}
& \multicolumn{4}{c}{Training Hours} \\
% \cmidrule(lr){3-6}
& & 2k & 4k & 6k & 8k \\
\midrule

\multicolumn{6}{l}{\textit{ASR Overall WER (\%) $\downarrow$}} \\
\midrule
\multirow{2}{*}{ASR}
& CE  & 113.89 & 9.73 & 7.00 & 6.89 \\
& LFD & \textbf{9.79} & \textbf{7.32} & \textbf{6.83} & \textbf{6.57} \\
\midrule

\multirow{2}{*}{QA}
& CE  & 123.91 & 11.82 & 7.89 & 8.23 \\
& LFD & \textbf{13.68} & \textbf{7.71} & \textbf{7.75} & \textbf{7.20} \\

\midrule
\multicolumn{6}{l}{\textit{QA Overall Accuracy (\%) $\uparrow$}} \\
\midrule
\multirow{2}{*}{ASR}
& CE  & 0.52 & 2.44 & 2.99 & 2.44 \\
& LFD & \textbf{35.90} & \textbf{40.04} & \textbf{40.13} & \textbf{40.20} \\
\midrule

\multirow{2}{*}{QA}
& CE  & 0.91 & 39.28 & \textbf{43.58} & 43.35 \\
& LFD & \textbf{38.11} & \textbf{43.28} & 43.48 & \textbf{43.93} \\
\bottomrule
\end{tabular}%
}
\end{table}

\begin{figure*}[t]
    \centering
    \begin{subfigure}[b]{0.48\textwidth}
        \centering
        \includegraphics[width=\linewidth]{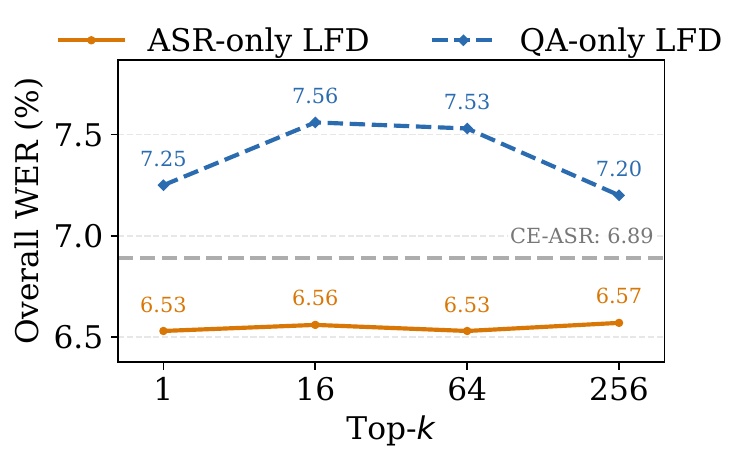}
        \caption{ASR performance (WER $\downarrow$).}
        \label{fig:topk_asr}
    \end{subfigure}
    \hfill
    \begin{subfigure}[b]{0.48\textwidth}
        \centering
        \includegraphics[width=\linewidth]{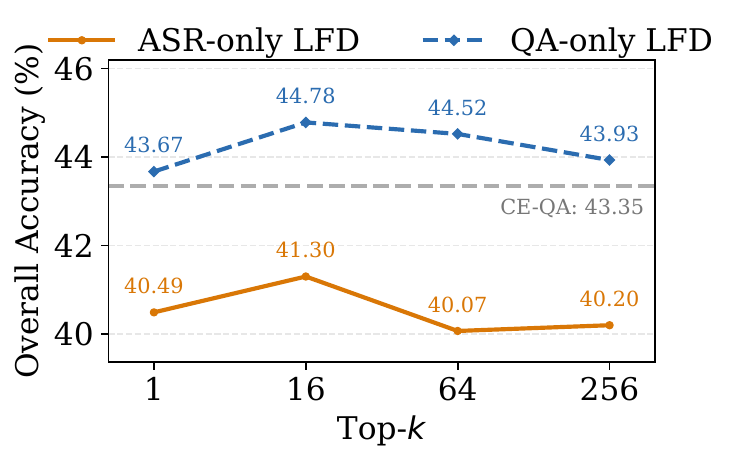}
        \caption{QA performance (ACC $\uparrow$).}
        \label{fig:topk_qa}
    \end{subfigure}
    \caption{Effect of distribution granularity (top-$k$) on ASR and QA performance under ASR-only and QA-only training settings. Dashed gray lines indicate the corresponding CE baselines.}
    \label{fig:topk}
\end{figure*}

\subsection{Scaling Behavior of CE and LFD}
\label{sec:scaling}

To investigate the scaling behavior of CE and LFD, we conduct ASR-only and QA-only training with 2k, 4k, 6k, and 8k hours of data, using a top-$k$ value of 256 for LFD. All experiments follow the same model architecture and training configurations described in Section~\ref{sec:experimental_setup}. Table~\ref{tab:scaling_overall} summarizes the overall ASR and QA performance, with detailed results provided in Appendix~\ref{sec:appendix_scaling}.

\textbf{LFD exhibits superior data efficiency.}
With limited data, LFD achieves substantially better ASR and QA performance than CE under both ASR-only and QA-only settings. While CE requires more data for effective speech--language alignment, LFD achieves strong cross-task performance at smaller scales. These results suggest that learning richer conditional output distributions provides more informative supervision, facilitating effective alignment with limited data.

%\textbf{LFD exhibits superior data efficiency.} With limited training data, LFD achieves substantially better ASR and QA performance than CE under both ASR-only and QA-only settings. While CE requires more data to establish effective speech--language alignment, LFD achieves strong cross-task performance even at smaller data scales. These results suggest that learning richer conditional output distributions provides more informative supervision, facilitating effective alignment with limited training data.

\textbf{Increasing data does not eliminate the task asymmetry of one-hot supervision.}
As training data increases, ASR-only CE substantially improves recognition performance but consistently fails to develop meaningful QA capability. In contrast, QA-only CE progressively improves both tasks, achieving strong QA performance while retaining reasonable ASR capability. This persistent asymmetry is consistent with our mapping perspective: the approximately one-to-one mapping of ASR, combined with one-hot supervision, repeatedly reinforces deterministic transcription targets, potentially favoring transcription-specific alignment. Importantly, increasing ASR data alone does not alleviate this limitation, suggesting that the observed capability degradation cannot be attributed solely to insufficient training data.

% \textbf{Increasing data does not eliminate the task asymmetry of one-hot supervision.} As training data increases, ASR-only CE substantially improves recognition performance but consistently fails to develop meaningful QA capability. In contrast, QA-only CE progressively improves both tasks, achieving strong QA performance while retaining reasonable ASR capability. This persistent asymmetry is consistent with our mapping perspective: the approximately one-to-one mapping of ASR, combined with one-hot supervision, repeatedly reinforces deterministic transcription targets, potentially favoring transcription-specific alignment. Importantly, increasing the amount of ASR data alone does not alleviate this limitation, suggesting that the observed capability degradation cannot be attributed solely to insufficient training data.

\textbf{Broadening supervision changes cross-task scaling behavior.}
Unlike ASR-only CE, ASR-only LFD develops strong QA capability even with limited training data, while continuously improving recognition performance as data increases. QA-only LFD also maintains strong performance on both tasks, with QA accuracy gradually approaching saturation. Although QA-only LFD retains a modest advantage in QA performance, the gap between the two training settings remains substantially smaller than under CE. These findings further support our proposed principle that broadening the learning objective beyond single-reference targets can mitigate task-asymmetric alignment effects, rather than relying solely on increasing the amount of task-specific data.

\begin{table}[t]
\centering
\caption{ASR and QA performance and token-level predictive entropy under CE and LFD supervision.}
\label{tab:entropy_analysis}
\renewcommand{\arraystretch}{1.05}
\setlength{\tabcolsep}{4pt}
\small
\resizebox{\columnwidth}{!}{%
\begin{tabular}{ccccccc}
\toprule
\multirow{2}{*}{System}
& \multirow{2}{*}{\begin{tabular}[c]{@{}c@{}}Training\\Strategy\end{tabular}}
& \multirow{2}{*}{\begin{tabular}[c]{@{}c@{}}Training\\Data\end{tabular}}
& \multicolumn{2}{c}{Performance}
& \multicolumn{2}{c}{Entropy} \\
\cmidrule(lr){4-5}
\cmidrule(lr){6-7}
& & & ASR $\downarrow$ & QA $\uparrow$ & ASR & QA \\
\midrule
A1 & \multirow{2}{*}{CE}
   & ASR & 6.89 & 2.44 & 0.25 & 0.25 \\
B1 & & QA & 8.23 & 43.35 & 0.17 & 0.39 \\
\midrule
A2 & \multirow{2}{*}{LFD}
   & ASR & 6.57 & 40.20 & 0.20 & 0.48 \\
B2 & & QA & 7.20 & 43.93 & 0.15 & 0.60 \\
\bottomrule
\end{tabular}%
}
\end{table}

\begin{table*}[t]
\centering
\caption{ASR and QA performance and token-level predictive entropy under full-parameter fine-tuning with CE and LFD supervision. ASR performance is reported as WER (\%), while QA performance is measured by accuracy (\%). C0 and D0 denote the corresponding adapter-only LFD initialization models.}
\label{tab:full_finetuning}
\renewcommand{\arraystretch}{1.05}
\centering
\resizebox{\textwidth}{!}
{%
\begin{tabular}{cccccccccc}
\toprule
\multirow{2}{*}{System} & \multirow{2}{*}{\begin{tabular}[c]{@{}c@{}}Training\\ Data\end{tabular}} & \multirow{2}{*}{\begin{tabular}[c]{@{}c@{}}Training\\ Strategy\end{tabular}} & \multicolumn{3}{c}{ASR(Librispeech $\downarrow$)} & \multicolumn{4}{c}{QA(TELEVAL $\uparrow$)}                        \\
                        &                                                                          &                                                                              & clean           & other          & Overall        & LlamaQA-en     & TriviaQA-en    & WebQ-en        & Overall        \\ \midrule
C0(init)                & \multirow{3}{*}{ASR}                                                     & LFD                                                                          & 4.45            & 8.62           & 6.53           & \textbf{72.33} & 32.50          & 39.01          & 40.49          \\
C1                      &                                                                          & CE                                                                           & \textbf{3.61}   & \textbf{7.43}  & \textbf{5.52}  & 0.33           & 2.39           & 1.91           & 1.89           \\
C2                      &                                                                          & LFD                                                                          & 3.70   & 7.46  & 5.57  & 72.00 & \textbf{32.86} & \textbf{42.78} & \textbf{42.93} \\ \midrule
D0(init)                & \multirow{3}{*}{QA}                                                      & LFD                                                                          & 5.24            & 9.27           & 7.25           & \textbf{73.33} & 36.68          & 42.11          & 43.67          \\
D1                      &                                                                          & CE                                                                           & 4.79            & 8.73           & 6.76           & 74.00          & 36.92          & 43.86          & 44.91          \\
D2                   &                                                                          & LFD                                                                          & \textbf{4.58}   & \textbf{8.38}  & \textbf{6.48}  & \textbf{73.33} & \textbf{38.23} & \textbf{44.07} & \textbf{45.33} \\ \bottomrule
\end{tabular}%
}
\end{table*}

\subsection{Distributional Analysis of ASR and QA}
\label{sec:distribution_analysis}

To investigate the task-dependent distributional characteristics of ASR and QA, we analyze the token-level predictive entropy of SLMs trained with CE and LFD, as reported in Table~\ref{tab:entropy_analysis}.

\textbf{Predictive entropy reflects task-specific input--output mappings.}
Except for A1, all systems exhibit lower predictive entropy on ASR than on QA, consistent with the distinct mapping characteristics discussed in Section~\ref{sec:mapping}. This difference reflects the relatively constrained output space of ASR and the diversity of valid responses in QA. In contrast, A1 exhibits similarly low entropy on both tasks and predominantly generates transcriptions even for QA inputs, suggesting a transcription-oriented bias induced by ASR-only CE supervision. Furthermore, as shown in Appendix~\ref{sec:appendix_entropy}, LFD-trained SLMs exhibit task-dependent entropy patterns similar to those of the pretrained text LLM, suggesting that distributional supervision helps preserve its task-conditioned generation characteristics.

\subsection{Effect of Distribution Granularity}
\label{sec:distribution_granularity}

To investigate the effect of distribution granularity, we vary top-$k$ over $\{1,16,64,256\}$ using 8,000 hours of training data under both ASR-only and QA-only settings. Fig.~\ref{fig:topk} shows the overall performance, with details in Appendix~\ref{sec:appendix_granularity}.
% To investigate the effect of distribution granularity, we vary the top-$k$ value over $\{1,16,64,256\}$ using 8,000 hours of training data under both ASR-only and QA-only settings. Fig.~\ref{fig:topk} presents the overall performance, with detailed results provided in Appendix~\ref{sec:appendix_granularity}.

\textbf{LFD is robust to distribution granularity.}
Across both training settings, varying top-$k$ has a relatively limited impact on ASR and QA performance, with even $k=1$ achieving competitive results. For ASR-only training, this may be attributed to the concentrated teacher distributions induced by the repetition instruction. For QA-only training, diverse generation trajectories in the training data may provide rich supervision across different contexts, reducing the need to explicitly preserve large numbers of token probabilities at each position.

\textbf{Sparse distribution supervision offers practical advantages.}
The competitive performance of $k=1$ suggests that distributional supervision need not require fine-grained token probabilities. Notably, our tail-aware objective retains the teacher's top-1 confidence and the remaining probability mass, rather than reducing supervision to a hard label. This finding suggests that sparse distribution supervision can substantially reduce the storage and data transfer overhead of teacher distributions, particularly in offline distillation.

% \textbf{Sparse distribution supervision offers practical advantages.} The competitive performance of $k=1$ suggests that effective distributional supervision does not necessarily require fine-grained token probabilities. Notably, our tail-aware objective retains the teacher's top-1 confidence and the remaining probability mass, rather than reducing supervision to a hard label. This finding suggests that sparse distribution supervision can substantially reduce the storage and data transfer overhead of teacher distributions, particularly in offline distillation.

\subsection{Behavior under Full-Parameter Fine-Tuning}
\label{sec:full_finetuning}

To investigate whether the task-asymmetric effects of CE persist beyond adapter-only alignment, we further conduct full-parameter fine-tuning experiments. For ASR-only training, both CE and LFD are initialized from the same ASR-only LFD model trained with top-$k=1$ during the alignment stage. Similarly, QA-only experiments are initialized from the corresponding QA-only LFD model. All training configurations are kept identical except for the supervision objectives and their associated settings, with top-$k=1$ used for LFD. Table~\ref{tab:full_finetuning} presents the results, including the performance of the initialization models.

\textbf{Task-asymmetric effects of one-hot supervision persist under full-parameter fine-tuning.}
From C0, ASR-only CE (C1) improves recognition performance but severely degrades QA capability. In contrast, QA-only CE (D1) improves both ASR and QA performance over its initialization (D0). This asymmetry is consistent with our mapping perspective: repeated optimization toward deterministic ASR targets under one-hot supervision may favor transcription-specific behavior, compromising previously acquired general-purpose capabilities. Importantly, this phenomenon persists even when the model starts with strong QA capabilities, suggesting that the conflict is not limited to the initial speech--language alignment stage.

% \textbf{Task-asymmetric effects of one-hot supervision persist under full-parameter fine-tuning.} Starting from C0, ASR-only CE (C1) improves recognition performance but severely degrades QA capability. In contrast, QA-only CE (D1) improves both ASR and QA performance over its initialization (D0). This asymmetry is consistent with our mapping perspective: repeated optimization toward deterministic ASR targets under one-hot supervision may favor transcription-specific behavior, compromising previously acquired general-purpose capabilities. Importantly, this phenomenon persists even when the model is initialized with strong QA capabilities, suggesting that the conflict is not limited to the initial speech--language alignment stage.

\textbf{Broadening ASR supervision mitigates capability degradation during full-parameter fine-tuning.}
Compared with C0, ASR-only LFD (C2) improves ASR and overall QA performance. Notably, C2 achieves nearly identical ASR performance to C1 while substantially outperforming it on all QA benchmarks. Similarly, QA-only LFD (D2) improves both ASR and overall QA performance over D0 and outperforms CE (D1) on both overall metrics. These findings support our principle that broadening supervision beyond single-reference targets can mitigate task-asymmetric effects while maintaining strong task performance, even during full-parameter fine-tuning.

% \textbf{Broadening ASR supervision mitigates capability degradation during full-parameter fine-tuning.} Compared with C0, ASR-only LFD (C2) further improves both ASR and overall QA performance. Notably, C2 achieves nearly identical ASR performance to C1 while substantially outperforming it on all QA benchmarks. Similarly, QA-only LFD (D2) improves both ASR and overall QA performance over D0 and outperforms CE (D1) on both overall metrics. These findings further support our proposed principle that broadening supervision beyond single-reference targets can mitigate task-asymmetric effects while maintaining strong task performance, even during full-parameter fine-tuning.

\begin{table}[t]
\centering
\caption{ASR and QA performance of CoT-based supervision. The base model is obtained through adapter-only alignment with mixed CoT-ASR and CoT-QA data. CoT-ASR and CoT-QA are then separately full-parameter fine-tuned from the same initialization.}
\label{tab:cot_asr}
\renewcommand{\arraystretch}{1.1}
\setlength{\tabcolsep}{4pt}
\small
\resizebox{\columnwidth}{!}{%
\begin{tabular}{llcc}
\toprule
Model & Fine-Tuning Data
& ASR WER (\%) $\downarrow$
& QA ACC (\%) $\uparrow$ \\
\midrule
Base Model & -- & 7.57 & \textbf{42.86} \\
CoT-ASR & ASR-only & \textbf{7.38} & 42.54 \\
CoT-QA & QA-only & 7.42 & 42.67 \\
\bottomrule
\end{tabular}%
}
\end{table}

\section{Beyond LFD: Rethinking ASR Supervision with CoT}
\label{sec:alternative_supervision}

The full-parameter fine-tuning results in Section~\ref{sec:full_finetuning} show that ASR-only CE training can severely degrade QA capabilities. To further examine our hypothesis regarding the interaction between ASR's deterministic mapping and single-reference supervision, we investigate Chain-of-Thought ASR (CoT-ASR) as an alternative strategy. Unlike direct transcription supervision, CoT-ASR broadens the generation space through intermediate reasoning while preserving transcription as the primary objective. We first obtain a base SLM through adapter-only alignment on mixed CoT-ASR and CoT-QA data, then separately perform full-parameter fine-tuning with CoT-ASR and CoT-QA data from this shared initialization.

As shown in Table~\ref{tab:cot_asr}, CoT-ASR improves recognition while largely preserving QA capability despite using only ASR data for full-parameter fine-tuning. CoT-QA similarly improves ASR while maintaining strong QA performance. These findings suggest that capability degradation is not inevitable under ASR-only training but may depend on the interaction between task mappings and supervision objectives. Thus, CoT-ASR provides an alternative realization of our principle of broadening ASR supervision beyond LFD.

% The full-parameter fine-tuning results in Section~\ref{sec:full_finetuning} show that ASR-only training with conventional CE supervision can severely degrade QA capabilities. To further examine our core hypothesis that this degradation is associated with the interaction between ASR's deterministic mapping and single-reference supervision, we investigate CoT-ASR as an alternative strategy consistent with our mapping perspective. Unlike direct transcription supervision, CoT-ASR introduces a broader generation space through intermediate reasoning while retaining transcription as the task-specific objective. We first obtain a base SLM through adapter-only alignment using mixed CoT-ASR and CoT-QA data. Starting from this shared base model, we separately perform full-parameter fine-tuning with CoT-ASR and CoT-QA data.

% As shown in Table~\ref{tab:cot_asr}, CoT-ASR improves recognition performance while largely preserving QA capability, despite using only ASR data during full-parameter fine-tuning. CoT-QA exhibits a similar trend, improving ASR performance while maintaining strong QA capability. These observations are consistent with our core position: the degradation of general-purpose capabilities is not an inevitable consequence of ASR-only training, but may depend on how its learning targets interact with task-specific mapping characteristics. By providing a broader learning space while preserving a clear ASR objective, CoT-ASR offers an alternative realization of our proposed supervision principle beyond LFD.

\section{Discussion}

\subsection{Task Mappings and Supervision as Joint Factors}

Our findings suggest that the ASR--QA conflict arises partly from the interaction between task-specific input--output mappings and supervision objectives. Although CE can learn conditional output distributions for both tasks, single-reference supervision may induce substantially different learning behaviors. In particular, ASR's relatively deterministic mapping, combined with repeated transcription supervision, may bias speech--language alignment toward ASR-specific behavior. This interpretation is consistent with persistent QA degradation under ASR-only CE despite increasing training data, as well as severe capability loss during full-parameter fine-tuning. The distinct entropy characteristics of ASR and QA further support our proposed mapping perspective.

% Our findings suggest that the conflict between ASR specialization and general-purpose capabilities should be understood through the interaction between task-specific input--output mappings and training supervision. Although CE can learn the conditional output distributions of both ASR and QA, its single-reference supervision may induce different learning behaviors across tasks. In particular, the relatively deterministic mapping of ASR, combined with repeated optimization toward transcription targets, may bias speech--language alignment toward ASR-specific behavior. This interpretation is consistent with the persistent QA degradation under ASR-only CE, even as training data increases, and the severe capability loss observed during full-parameter fine-tuning. The distinct entropy characteristics of ASR and QA further reflect their task-dependent output spaces, providing additional support for our mapping perspective.

Our results also highlight broadening ASR supervision as a promising solution without changing the recognition objective. LFD realizes this principle by learning the pretrained LLM's conditional output distributions beyond individual reference tokens, while CoT-ASR broadens the generation space through intermediate reasoning. Both strategies preserve ASR as the primary objective, suggesting that general-purpose capability degradation is not inevitable under ASR-only training but depends partly on how task-specific data is supervised during training.

% More importantly, our results suggest that broadening ASR supervision offers a promising direction for mitigating this conflict without changing the underlying recognition objective. LFD realizes this principle by preserving the pretrained LLM's conditional output distributions, enabling the model to learn beyond individual reference tokens. CoT-ASR provides an alternative realization by introducing a broader generation space through intermediate reasoning while retaining transcription as the task objective. The effectiveness of these different strategies suggests that severe general-purpose capability degradation is not an inevitable consequence of ASR-only training, but is closely related to how the model learns from task-specific data.

\subsection{Relation to Prior Distribution-based Alignment Methods}

Prior studies have explored different perspectives to address the degradation of general-purpose capabilities in SLMs. DiVA~\cite{Diva} and BLSP-KD~\cite{Blsp-kd} leverage cross-modal distillation to transfer pretrained LLM capabilities to speech, while SALAD~\cite{Salad} investigates capability degradation from the perspectives of catastrophic forgetting, cross-modal misalignment, and domain mismatch. Although these studies propose effective approaches to preserving general-purpose capabilities, they do not systematically examine why ASR-oriented training can lead to particularly severe capability degradation.

In contrast, our work introduces a mapping-based perspective to investigate the interaction between task-specific input--output mappings and one-hot CE supervision. We argue that the relatively deterministic mapping of ASR, combined with single-reference supervision, may bias models toward transcription-specific behavior, leading to asymmetric effects on ASR and QA capabilities. Based on this perspective, we advocate broadening the learning space of ASR while preserving its recognition objective. LFD realizes this principle by learning the conditional output distributions of pretrained LLMs, while CoT-ASR provides an alternative realization through richer generation trajectories. Our contribution lies not only in the method but also in a new perspective for understanding and mitigating the ASR--QA conflict.
% In contrast, our work introduces a mapping-based perspective to investigate the interaction between task-specific input--output mappings and one-hot CE supervision. We argue that the relatively deterministic mapping of ASR, combined with single-reference supervision, may bias models toward transcription-specific behavior, leading to asymmetric effects on ASR and QA capabilities. Based on this perspective, we advocate broadening the learning space of ASR while preserving its recognition objective. LFD realizes this principle by learning the conditional output distributions of pretrained LLMs, while CoT-ASR provides an alternative realization through richer generation trajectories. Our contribution therefore lies not only in the proposed method, but also in a new perspective for understanding and mitigating the ASR--QA conflict.

\section{Conclusion}

In this work, we revisit the ASR--QA conflict in SLMs through the interaction between task-specific input--output mappings and supervision objectives. We reveal task-asymmetric effects of one-hot supervision and advocate broadening ASR learning objectives while preserving recognition accuracy. We realize this principle through Learning from Distribution (LFD), which leverages the conditional output distributions of pretrained LLMs to provide richer supervision. Experiments demonstrate that LFD improves both ASR and QA, mitigates task asymmetry, and preserves cross-task capabilities during full-parameter fine-tuning. CoT-ASR further supports the broader applicability of this principle, highlighting the importance of mapping-aware supervision for general-purpose SLMs.

\section{Limitations}

Our experiments are conducted on a single SLM architecture based on Whisper-large-v3 and Qwen3-4B. Although our mapping-based perspective focuses on task characteristics and supervision objectives rather than specific model architectures, its generalizability across different model families and scales remains to be explored.

Furthermore, our analysis focuses on ASR and spoken QA, leaving its applicability to other speech-language tasks for future investigation. Extending this perspective to tasks with diverse input--output mappings may provide further insights into the interaction between task characteristics and supervision objectives.

% \section*{Acknowledgments}

% Bibliography entries for the entire Anthology, followed by custom entries
%\bibliography{custom,anthology-overleaf-1,anthology-overleaf-2}

% Custom bibliography entries only
\bibliography{custom}

\appendix

\section{Task-Conditioned Entropy Analysis}
\label{sec:appendix_entropy}

\begin{table}[h]
\centering
\caption{Token-level predictive entropy of text LLMs and SLMs on ASR and QA benchmarks. Lower values indicate more concentrated predictive distributions. LFD-ASR denotes the model obtained through full-parameter fine-tuning in Section~\ref{sec:full_finetuning}.}
\label{tab:appendix_entropy}
\renewcommand{\arraystretch}{1.1}
\setlength{\tabcolsep}{3.5pt}
\small
\resizebox{\columnwidth}{!}{%
\begin{tabular}{llccccc}
\toprule
\multirow{2}{*}{Model}
& \multirow{2}{*}{Type}
& \multicolumn{2}{c}{ASR}
& \multicolumn{3}{c}{QA} \\
\cmidrule(lr){3-4}
\cmidrule(lr){5-7}
& & test-clean & test-other
& LlamaQA-en & TriviaQA-en & WebQ-en \\
\midrule
Qwen3-4B & LLM & 0.04 & 0.04 & 0.58 & 0.67 & 0.58 \\
LFD-ASR & SLM & 0.10 & 0.17 & 0.59 & 0.72 & 0.62 \\
LFD-QA  & SLM & 0.09 & 0.13 & 0.54 & 0.66 & 0.56 \\
\bottomrule
\end{tabular}%
}
\end{table}

To further investigate the distributional characteristics of ASR and QA, we analyze the token-level predictive entropy of both text LLMs and SLMs. Given a model's predictive distribution $p_t$ over vocabulary $\mathcal{V}$ at generation step $t$, the entropy is defined as
\begin{equation}
    H(p_t) = -\sum_{v \in \mathcal{V}}
    p_t(v)\log p_t(v).
\end{equation}
We report the average predictive entropy across response tokens for each evaluation benchmark. Text LLMs are conditioned on speech transcripts, whereas SLMs receive the corresponding speech inputs.

As shown in Table~\ref{tab:appendix_entropy}, both text LLMs and SLMs consistently exhibit substantially lower predictive entropy on ASR than on QA. This pattern is consistent with their distinct input--output mappings: ASR typically involves relatively deterministic transcription targets, whereas QA allows multiple semantically valid responses.

Notably, the pretrained Qwen3-4B exhibits the same task-dependent entropy pattern, suggesting that these distributional differences are not specific to speech--language alignment. Furthermore, the SLMs retain this distinction, exhibiting concentrated predictions on ASR and more diverse predictive distributions on QA. These observations provide additional evidence that the entropy characteristics of ASR and QA reflect their underlying task-specific mapping structures.

\begin{table*}[!t]
\centering

% ==========================
% Table 1: Scaling Results
% ==========================

\caption{Detailed scaling results of CE and LFD under ASR-only and QA-only training settings with 2,000 to 8,000 hours of training data. ASR performance is reported as WER (\%), while QA performance is measured by accuracy (\%) on TELEVAL. Lower WER and higher accuracy indicate better performance.}
\label{tab:scaling_results}

\renewcommand{\arraystretch}{1.05}
\resizebox{\textwidth}{!}{%
\begin{tabular}{c|c|c|ccc|cccc}
\hline
\multirow{3}{*}{\begin{tabular}[c]{@{}c@{}}Training\\Data\end{tabular}}
& \multirow{3}{*}{\begin{tabular}[c]{@{}c@{}}Training\\Strategy\end{tabular}}
& \multirow{3}{*}{\begin{tabular}[c]{@{}c@{}}Data\\Duration\end{tabular}}
& \multicolumn{3}{c|}{ASR}
& \multicolumn{4}{c}{QA (TELEVAL $\uparrow$)} \\

& & & \multicolumn{3}{c|}{LibriSpeech-test $\downarrow$}
& \multirow{2}{*}{LlamaQA-en}
& \multirow{2}{*}{TriviaQA-en}
& \multirow{2}{*}{WebQ-en}
& \multirow{2}{*}{Overall} \\

& & & clean & other & Overall & & & & \\
\hline

\multirow{8}{*}{ASR}
& \multirow{4}{*}{CE}
& 2000 & 113.17 & 114.62 & 113.89 & 0.52 & 0.96 & 0.10 & 0.52 \\
& & 4000 & 7.31 & 12.16 & 9.73 & 2.33 & 2.75 & 2.32 & 2.44 \\
& & 6000 & 4.83 & 9.17 & 7.00 & 2.67 & 2.51 & 3.25 & 2.99 \\
& & 8000 & 4.83 & 8.96 & 6.89 & 1.67 & 2.87 & 2.37 & 2.44 \\
\cline{2-10}

& \multirow{4}{*}{LFD}
& 2000 & 7.38 & 12.21 & 9.79 & 69.33 & 30.47 & 33.08 & 35.90 \\
& & 4000 & 5.12 & 9.54 & 7.32 & 68.67 & 31.66 & 39.83 & 40.04 \\
& & 6000 & 4.71 & 8.95 & 6.83 & 68.33 & 32.62 & 39.01 & 40.13 \\
& & 8000 & 4.51 & 8.63 & 6.57 & 65.33 & 31.78 & 39.94 & 40.20 \\
\hline

\multirow{8}{*}{QA}
& \multirow{4}{*}{CE}
& 2000 & 121.52 & 126.31 & 123.91 & 1.00 & 2.03 & 0.41 & 0.91 \\
& & 4000 & 8.88 & 14.78 & 11.82 & 70.00 & 31.54 & 37.87 & 39.28 \\
& & 6000 & 5.63 & 10.15 & 7.89 & 74.33 & 35.01 & 42.52 & 43.58 \\
& & 8000 & 6.35 & 10.11 & 8.23 & 72.00 & 36.08 & 42.05 & 43.35 \\
\cline{2-10}

& \multirow{4}{*}{LFD}
& 2000 & 10.63 & 16.75 & 13.68 & 69.00 & 29.27 & 37.15 & 38.11 \\
& & 4000 & 5.56 & 9.88 & 7.71 & 73.67 & 37.04 & 41.28 & 43.28 \\
& & 6000 & 5.33 & 10.19 & 7.75 & 72.67 & 36.08 & 42.16 & 43.48 \\
& & 8000 & 5.18 & 9.23 & 7.20 & 73.33 & 36.56 & 42.57 & 43.93 \\
\hline
\end{tabular}%
}

\vspace{1.2em}

% ==========================
% Table 2: Top-k Results
% ==========================

\caption{Detailed results of LFD with different top-$k$ values under ASR-only and QA-only training settings. ASR performance is reported as WER (\%) on LibriSpeech, while QA performance is measured by accuracy (\%) on TELEVAL. Lower WER and higher accuracy indicate better performance.}
\label{tab:topk_results}

\renewcommand{\arraystretch}{1.1}
\resizebox{0.92\textwidth}{!}{%
\begin{tabular}{c|c|ccc|cccc}
\hline
\multirow{2}{*}{\begin{tabular}[c]{@{}c@{}}Training\\Data\end{tabular}}
& \multirow{2}{*}{Top-$k$}
& \multicolumn{3}{c|}{ASR (LibriSpeech-test $\downarrow$)}
& \multicolumn{4}{c}{QA (TELEVAL $\uparrow$)} \\

& & clean & other & Overall
& LlamaQA-en & TriviaQA-en & WebQ-en & Overall \\
\hline

\multirow{4}{*}{ASR}
& 1   & 4.45 & 8.62 & 6.53 & 72.33 & 32.50 & 39.01 & 40.49 \\
& 16  & 4.47 & 8.65 & 6.56 & 71.67 & 32.74 & 40.30 & 41.30 \\
& 64  & 4.51 & 8.56 & 6.53 & 69.00 & 32.86 & 38.70 & 40.07 \\
& 256 & 4.51 & 8.63 & 6.57 & 65.33 & 31.78 & 39.94 & 40.20 \\
\hline

\multirow{4}{*}{QA}
& 1   & 5.24 & 9.27 & 7.25 & 73.33 & 36.68 & 42.11 & 43.67 \\
& 16  & 5.20 & 9.92 & 7.56 & 74.33 & 36.92 & 43.60 & 44.78 \\
& 64  & 5.17 & 9.90 & 7.53 & 74.00 & 37.28 & 43.09 & 44.52 \\
& 256 & 5.18 & 9.23 & 7.20 & 73.33 & 36.56 & 42.57 & 43.93 \\
\hline
\end{tabular}%
}

\end{table*}

\section{Evaluation Benchmarks and Metrics}
\label{sec:appendix_evaluation}

\textbf{ASR Evaluation.} We evaluate ASR performance on LibriSpeech~\cite{librispeech}, a widely used English speech recognition benchmark derived from public-domain audiobooks. We use the test-clean and test-other subsets, which represent different recognition difficulty levels. Performance is measured by word error rate (WER), defined as $(S+D+I)/N$, where $S$, $D$, and $I$ denote substitution, deletion, and insertion errors, respectively, and $N$ is the number of reference words. Overall WER is averaged across the two subsets.

\textbf{Spoken QA Evaluation.} We evaluate spoken QA using three English subsets of TELEVAL~\cite{televal}: LlamaQA-en (300 samples), TriviaQA-en (837 samples), and WebQ-en (1,938 samples). These subsets assess spoken question answering across general knowledge and factoid QA tasks. Following TELEVAL's evaluation protocol, responses are evaluated against multiple acceptable reference answers using string matching. We report answer accuracy (ACC) for each subset, with overall accuracy computed across all 3,075 samples.

\section{Detailed Scaling Results}
\label{sec:appendix_scaling}

Table~\ref{tab:scaling_results} presents detailed scaling results of CE and LFD under ASR-only and QA-only training settings. We evaluate both supervision objectives with 2,000, 4,000, 6,000, and 8,000 hours of training data to investigate their effectiveness across different data scales. ASR performance is reported as WER on LibriSpeech test-clean and test-other, while QA performance is measured by accuracy on three English TELEVAL subsets. Individual benchmark results and overall performance are provided for comprehensive comparison.

The results show that LFD provides substantial advantages over CE at smaller data scales. As training data increases, ASR-only CE improves recognition performance but consistently exhibits limited QA capability. In contrast, LFD achieves stronger cross-task performance under both training settings. These observations further support the data efficiency and task-asymmetric learning effects discussed in Section~\ref{sec:scaling}.

\section{Detailed Distribution Granularity Results}
\label{sec:appendix_granularity}

Table~\ref{tab:topk_results} presents detailed results of LFD with different top-$k$ values under ASR-only and QA-only training settings. We investigate distribution granularity using $k \in \{1,16,64,256\}$, while keeping all other training configurations unchanged. For each setting, we report ASR performance on LibriSpeech test-clean and test-other, along with QA accuracy on three English TELEVAL subsets.

Across different top-$k$ values, LFD maintains relatively stable performance on both ASR and QA tasks, suggesting limited sensitivity to distribution granularity. Notably, even $k=1$ achieves competitive results when the residual probability mass is preserved through tail-aware supervision. This finding highlights the practicality of compact teacher distribution storage for efficient offline training. Further analysis is provided in Section~\ref{sec:distribution_granularity}.

% \section{Detailed Scaling Results}
% \label{sec:appendix_scaling}

% Table~\ref{tab:scaling_results} presents the detailed scaling results of CE and LFD under ASR-only and QA-only training settings. We evaluate both supervision objectives using 2,000, 4,000, 6,000, and 8,000 hours of training data. Results are reported on LibriSpeech and TELEVAL, including individual benchmarks and overall performance. The corresponding analysis is provided in Section~\ref{sec:scaling}.

% \section{Detailed Distribution Granularity Results}
% \label{sec:appendix_granularity}

% Table~\ref{tab:topk_results} presents the detailed results of LFD with different top-$k$ values under ASR-only and QA-only training settings. We evaluate distribution granularity using $k \in \{1,16,64,256\}$, while keeping all other training configurations unchanged. Results are reported on LibriSpeech and TELEVAL, including individual benchmarks and overall performance. The corresponding analysis is provided in Section~\ref{sec:distribution_granularity}.

\end{document}